\documentclass{aa}

\usepackage{graphics}

\begin{document}
\thesaurus{08 (08.09.2 BD+31$\degr$643; 08.03.4; 08.16.2; 
03.13.4; 10.15.2 IC~348)}

\title{Dust distribution in radiation pressure outflow. \\
Application to the BD+31$\degr$643 disk}

\author{
A. Lecavelier des Etangs \inst{1,2}
\and A. Vidal-Madjar \inst{2}
\and R. Ferlet \inst{2}
}
        
\offprints{A. Lecavelier des Etangs}
\institute{
NCRA, TATA Institute of Fundamental Research, Post Bag 3, Ganeshkhind, 
Pune University Campus, Pune 411 007, India
\and Institut d'Astrophysique de Paris, CNRS, 98bis Boulevard Arago, F-75014 Paris, France
}

\date{Received / Accepted}

\titlerunning{Dust distribution in the BD+31$\degr$643 disk}

\maketitle

\begin{abstract}

The radial distribution of dust has been calculated in outflows driven
by radiation pressure in which the gas drag is taken into account.
It is shown that the dust distribution in the newly discovered 
candidate disk
around BD+31$\degr$643\ can be explained by this simple model, provided that
the ambient gas of the surrounding cluster IC~348 has a density 
$\rho\sim 10^4~$cm$^{-3}$. The dust is probably produced 
within 2300~AU from the central binary star.

\keywords{stars: \mbox{BD+31$\degr$643} -- circumstellar
matter -- planetary systems }

\end{abstract}

\section{Introduction}
\label{Intro}

Dust around main sequence stars has been discovered more than a decade
ago by the IRAS satellite (Aumann et al. 1984), but
little work on the dust dynamics has been done. 
Moreover until recently, $\beta\:$Pictoris was the only star to 
capture all the attention, because images showed that the dust 
shell was in fact
a disk seen edge-on from the Earth (Smith \& Terrile 1984). These
images have given unique information 
on disk morphology and dust distribution (Artymowicz et al. 1989, 
Kalas \& Jewitt 1995). The dust distribution, 
the intriguing asymmetries or the wrapped geometry
have been studied thoroughly (Roques et al. 1994, Artymowicz 1995, 
Lecavelier des Etangs et al. 1996a, 1996b, Mouillet et al. 1997).
But it is still not clear if these results apply only to
the well-known $\beta\:$Pictoris\ disk or if they describe a general
scheme for which $\beta\:$Pictoris\ is an example.

Recently, the situation changed drastically 
when Kalas \& Jewitt (1997) discovered a candidate dust disk 
around BD+31$\degr$\-643, although it is still to be confirmed
by observational follow-up. 
As in the $\beta\:$Pictoris\ case, the dust lifetime around 
BD+31$\degr$\-643\ is smaller than the age
of the star. There must therefore be a replenishment mechanism. 
The origin of dust (collision or evaporation of parent bodies?) 
must be identified.
It is interesting to see if the analysis of the dynamics and the distribution
of the dust can provide new insight on various aspects of the origin
and history of dusty disks around main-sequence stars.

If this dust was in fact coming from the interstellar
medium and dragged by invisible massive gas, the 
dynamics of this disk could be very different from the
dynamics of the dust around $\beta\:$Pictoris. But, if this disk was really
the second member of the $\beta\:$Pictoris\ family for which we have an image,
then this would shed more light on 
the origin of the $\beta\:$Pictoris-like disks (Lissauer 1997).

In addition, BD+31$\degr$643\ is a binary star; this could have consequences
on possible dust production mechanisms. At least, if 
this disk is replenished like the $\beta\:$Pictoris one,  
this shows that the formation of planetesimals is possible around
binary stars (Kalas \& Jewitt, 1997).
The evolution and stability of planetary systems in this 
environment is starting to be investigated (Brunini 1997,
Benest 1998). But knowledge on these putative parent bodies
from the observation of dust have to be improved,
and a better understanding of the dust dynamics is needed.

In this paper, after a summary of the characteristics of the 
BD+31$\degr$643\ candidate disk,
where it is recalled that the dust is expelled by the radiation pressure
(Sect.~\ref{The bd disk}), we present theoretical calculation 
on outflows driven
by radiation pressure in which the gas drag is taken into account
(Sect.~\ref{Theoretical considerations} and~\ref{Outflow with gas drag}).
Application to BD+31$\degr$643\ (Sect.~\ref{appli. The bd disk}) 
and numerical simulations (Sect.~\ref{Numerical results})
show that the observed dust distribution (or more precisely 
the surface brightness distribution along the midplane of the disk) 
can be explained by the ambient gas drag of the cluster IC~348, 
if the gas density is between $10^3$ and $10^5$~cm$^{-3}$.

\section{The BD+31$\degr$643\ disk}
\label{The bd disk}

The apparent dust surrounding BD+31$\degr$643 presents strong 
similarities with the $\beta\:$Pictoris\ circumstellar matter,
namely a disk-like structure seen edge-on with  
a scattered light distribution close to a power law.
The similarities are clearly visible through a 
comparison of the morphology of both disks 
seen on the corresponding images 
(see Fig.~1 in Kalas \& Jewitt, 1997).

But important differences are also found.
The observed dust is at significantly larger distances in the case of 
BD+31$\degr$643, 
simply because the central star is much farther ($\sim 330$~pc).
In the surrounding of BD+31$\degr$643, the disk is seen between 1300~AU and 
6000~AU with a flatter brightness distribution: $F(r)\propto r^{-0.3}$
between 4 and 7~arcsec (1300--2300~AU), and 
$F(r)\propto r^{-1.9}$
between 7 and 20~arcsec (2300--6000~AU).

Most importantly, the large albedo of the dust in the blue shows that 
this particular dust is smaller than the wavelength with 
a size $\sim 0.1\mu$m. As mentioned by Kalas \& Jewitt,
these small grains must be more sensitive to the radiation
pressure than to the star gravity 
and must then be flowing out from the system. 
Indeed, the central star is a binary system more massive and more
luminous than $\beta\:$Pictoris\ (2$\times$5M$_{\odot}$, 2$\times$830L$_{\odot}$); 
the ratio of the stellar luminosity to the stellar mass is then 35
times larger for BD+31$\degr$643\ than for $\beta\:$Pictoris. 
Particles of size $\sim 0.1\mu$m have a $\beta$ ratio
of radiation force to gravitational force typically
3 to 10 times larger than particles of 1$\mu$m size (Burns et al., 1979;
Artymowicz 1988).
If we simply scale the results of Burns et al. (1979) for 
the solar system and of Artymowicz (1988) for $\beta$~Pic to the 
mass-luminosity ratio of BD+31$\degr$643, we find that 
particles around BD+31$\degr$643\ have $\beta$ ratios between 
100 and 350.

\begin{figure}
\resizebox{\hsize}{!}{\includegraphics{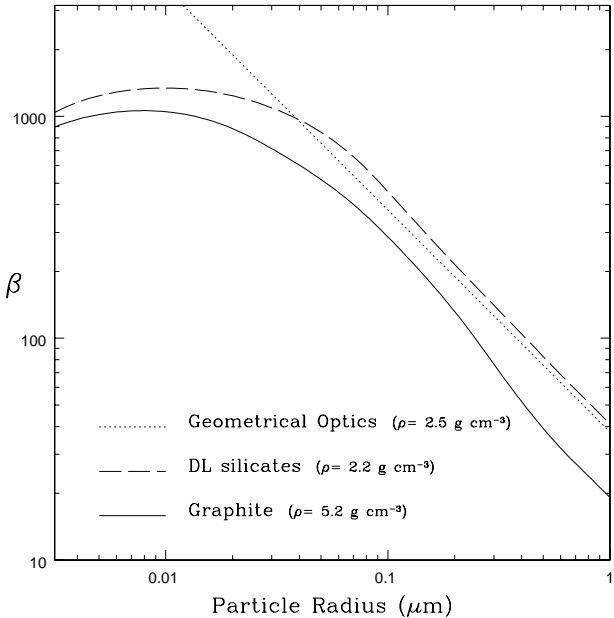}}
\caption[]{The ratio of the radiation force to the gravitational force
as a function of particle size for ``Draine \& Lee'' silicates (dashed
line), graphite (solid line) and the geometrical optics (dotted
line)}
\label{b_s}
\end{figure}

In fact, the $\beta$ ratio could even reach $\sim$ 1000 due to the large
flux in the UV from a B5V star.
With the optical properties given by Draine \& Lee (1984) for silicates 
(``DL silicates'') and for graphite, we carried out the $\beta$ ratio 
as a function of the size of particles (Fig~\ref{b_s}).
We have taken the UV spectrum measured by IUE on two B5V stars
used as references (HD~34739 and HD~199081).
The radiation pressure coefficient averaged over the stellar 
spectrum $<Q_{\rm pr}>$ has a major contribution from the
flux below 1500\AA\ where the absorption efficiency barely depends
on the particle size for particles larger than 0.01$\mu$m.
Consequently, we find that, in contrast to the solar system case, the $\beta$ ratio 
does not reach its maximum at about 0.1$\mu$m but around 0.01$\mu$m. 
This shows that 
particles seen around BD+31$\degr$643\
must have $\beta$ ratio larger than $\sim$100 
and must be ejected on hyperbolic orbits by radiation pressure.

Gas is also present within the young cluster IC~348 
in which BD+31$\degr$643\ is embedded.
Its presence is deduced from spectroscopic observations of
absorption lines and from radio maps (Snow et al. 1994). Snow et al.
showed that the column density of gas must be 
$N_{HI}\approx 3\cdot 10^{21}$~cm$^{-2}$, which is also consistent with the
extinction of the star ($E_{B-V}$=0.84, $A_V=2.8$). 
This allow to obtain observational limits on the gas density in the
circumstellar environment. With an observed maximal size of $\sim 1$~pc for
the IC~348 cluster, the gas density has a lower limit
$\rho_l \sim 10^3$cm$^{-3}$.
The spectroscopic observations of Snow et al. 
show that BD+31$\degr$643\ is inside the cloud. It
is also unlikely that the gas density in the star environment
is smaller than the mean value in the cluster,
because BD+31$\degr$643\ is at the center of the cluster
and coincides with the peak of dust extension and gaseous 
radio emission map. 
On the other hand, if the whole column density
of the gas was concentrated into the observed disk of size $>2000$~AU,
we would obtain an upper limit of the gas density $\rho_u \sim 10^5$cm$^{-3}$.

In short, the observations show that the disk 
of BD+31$\degr$643 has a particular environment with a gaseous medium 
which density is between $10^3$cm$^{-3}$ and $10^5$cm$^{-3}$.

\section{Outflow}

\label{Theoretical considerations}

\begin{figure*}
\resizebox{\hsize}{!}{\includegraphics{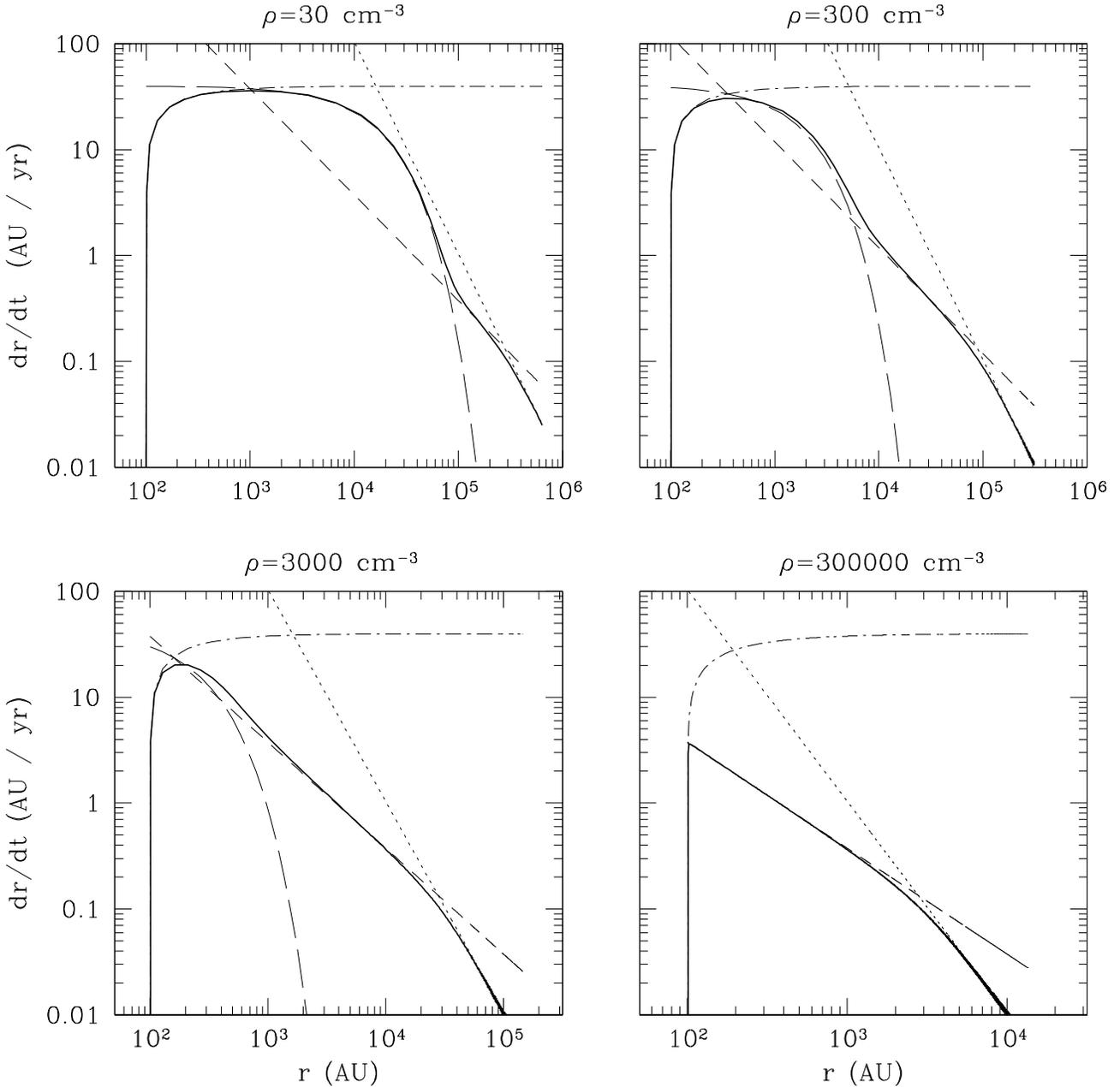}}
\caption[]{Plot of the dust radial velocity ($dr/dt$) as a function of the 
distance
to the star ($r$) for different gas density ($\rho$). 
The dust is supposed to be 
produced at a distance $r_0=100$~AU from the star, the ratio 
of the radiation force to the gravitational force is $\beta=200$.
The full Eq.~\ref{eq1} has been numerically solved (solid line).
The fit to the different phases: acceleration (dot-dashed lines), deceleration
(long-dashed line), stationary outflow (short-dashed and dotted lines)
are calculated with the approximation given by the equations of
Sect~\ref{Outflow with gas drag}.\\
For $\rho = 30 $cm$^{-3}$, the particle is accelerated between 100 and
1000~AU; it is then decelerated between 1000 and $10^5$~AU and reaches the
stationary outflow beyond that point.\\
For $\rho = 300 $cm$^{-3}$, the acceleration stops at 300~AU, and
the deceleration takes places between 300 and 5000~AU; the stationary outflow 
starts at $r_2=5000$~AU. The distribution changes again at $r_3=10^5$~AU when 
the thermal drag becomes dominant. \\
For $\rho = 3000 $cm$^{-3}$, the deceleration phase is very short
(between $r_1=150$~AU and $r_2=400$~AU). 
The stationary outflow gives a
radial velocity $\dot{r} \propto r^{-1}$
(between $r_2=400$~AU and $r_3=$30\,000~AU). 
The velocity regime changes at 30\,000~AU beyond which we have 
$\dot{r} \propto r^{-2}$.\\
For $\rho = 300\,000 $cm$^{-3}$, there is no deceleration phase at all.
The thermal drag is dominant when $r>r_3\cong 2800$~AU. 
}
\label{velocity}
\end{figure*}

When a dust particle with $\beta\gg 1$ is produced, it is ejected from the 
system by radiation pressure. It reaches rapidly a constant  
asymptotic velocity $v_{\infty} \sim v_0 \sqrt{2\beta-1}$,
where $v_0$ is the initial velocity of the parent body on its keplerian
orbit.
Consequently, if 
some dust is produced close to the star, it is spread in a disk-like
structure with a radial density $n \propto r^{-2}$. 
Nakano (1990) has given the first order relation between the radial
distribution of the dust density $n(r)$ 
and $F$ the apparent surface brightness along the midplane 
of the disk if seen edge-on from the Earth: $F \propto n(r) r^{-1} $.
As a result, the disk produced by outflowing dust has a 
radial brightness  
$F \propto r^{-3}$. 

But, as already noted in Lecavelier des Etangs et al. (1996a), 
the gas drag in an edge-on disk decreases the gradient of the observed 
brightness.
The next section is dedicated to the evaluation of 
the dust distribution taking into account the gas drag.

\section{Outflow with gas drag}
\label{Outflow with gas drag}

The motion of one particle driven by the radiation pressure
and dragged by the ambient gas is
\begin{equation}
\frac{d\vec{v}}{dt}=(\beta-1)\frac{GM}{r^3} \vec{r} + \vec{F_{PR}} + \vec{F_g} 
\label{eq1}
\end{equation}
where $\vec{r}$ and $\vec{v}$ are the position 
and the velocity of the particle, $M$ is the mass of the central star,
$\vec{F_{PR}}$ and $\vec{F_g}$ are the Poynting-Robertson and the
gas drags. The Poynting-Robertson effect 
is negligible compared to the gas drag as soon as the gas density is
larger than $\sim 1$cm$^{-3}$. The gas drag is given by (Kwok 1975):
\begin{equation}
 \vec{F_g} =  - (\sigma /m) m_\mathrm{H} \mu \rho 
                                         \sqrt{v_T^2 + (\vec{v}-\vec{v_g})^2 }
                (\vec{v}-\vec{v_g})
\label{Fg}
\end{equation}
where $\sigma$ and $m$ are the dust cross section and mass,
$\mu$ is the mean molecular weight of the gas and
$m_\mathrm{H}$ is the hydrogen mass.
$v_T$ is the sound velocity.
$\vec{v_g}$ is the gas velocity and is very small, 
we have taken $\vec{v_g}=0$.

For dust grains, we have $\sigma /m \cong 3/(4s\rho_d)$ where
$s$ and $\rho_d$ are the dust size and density.
We have shown that $s\propto \beta^{-1}$, where $s\beta\sim 20\mu$m
(Sect~\ref{The bd disk}).
Then, with $\rho_d\approx 1$~g~cm$^{-3}$, we obtain
$\sigma /m \approx 375 \beta $~cm$^2$~g$^{-1}$.

From the IRAS 60$\mu$m emission we know that the star environment
must be at a very low temperature: we can assume $T\sim 30$~K
(Snow et la. 1994). 
Then, as soon as the particle velocity is larger than 
$v_T= 0.14 \sqrt{(T/30 {\mathrm K})/\mu}$~AU~year$^{-1}$,
the first term under the square root of Eq.~\ref{Fg} is negligible,
and assuming $\mu =1$, we get
\begin{equation}
 \vec{F_g} \approx - 375 \beta m_\mathrm{H} \mu  \rho |(\vec{v}-\vec{v_g})|
                (\vec{v}-\vec{v_g})
        \approx - C \beta \rho |\vec{v}|\vec{v}
\label{Fg2}
\end{equation}
where $C=9.39\cdot 10^{-9}$cm$^{3}$AU$^{-1}$.\\

In order to evaluate the radial distribution of dust, we carry out
the different phase of the motion by approximate solutions of Eq.~\ref{eq1}.
We consider only the radial component in the polar coordinates
$(r,\theta)$ of the vectorial Eq.~\ref{eq1}:
\begin{equation}
\frac{d^2r}{dt^2} =r(\frac{d\theta}{dt})^2 + 
                   (\beta-1) \frac{GM}{r^2} - C\beta\rho(\frac{dr}{dt})^2
\label{eq1.1}
\end{equation}

Since we know that 
$n(r)\sim\dot{r}^{-1}r^{-2}$, we simply need to determine the radial
velocity $\dot{r}$ as a function of the distance to the star, $r$. Then,
the apparent surface brightness along the 
midplane of the disk, $F \propto n(r) r^{-1} $, 
can be compared to the observations.
If $\beta\gg 1$, the Eq.~\ref{eq1.1} becomes
\begin{equation}
\frac{d^2r}{dt^2}   \approx \beta \frac{GM}{r^2} - C\beta\rho(\frac{dr}{dt})^2
\label{eq2}
\end{equation}

\subsection{Acceleration phase}

When a particle is produced, its radial distance and velocity are small. 
Consequently, the last term of Eq.~\ref{eq2} is negligible and we 
obtain
\begin{equation}
\frac{d^2r}{dt^2}   \approx \beta \frac{GM}{r^2}  .
\label{acc}
\end{equation}
This gives $\dot{r}= \sqrt{2\beta GM}  \sqrt{r_0^{-1}-r^{-1}}$,
where $r_0$ is the initial radial distance at which the dust was produced.
This approximation is plotted on Fig.~\ref{velocity} with dot-dashed
lines. This is valid for small velocity and small distances, as long as
$d^2r/dt^2 > 0$, that is to say for $r<r_1 $ where 
\begin{equation}
r_1 \equiv \left(r0+\sqrt{r_0^2+2r_0/C\beta\rho}\right)/2  .
\end{equation}

\subsection{Deceleration phase}

After the particle has gone farther from the star, where the
radiation pressure becomes less important, and has gained a large radial
velocity, the gas drag becomes dominant and we have
\begin{equation}
\frac{d^2r}{dt^2} \approx -\left( C\beta\rho  
                                      - \beta\frac{GM}{r^2 \dot{r}^2 } \right)
                           \left( \frac{dr}{dt} \right)^2
\approx -\alpha \left( \frac{dr}{dt} \right)^2.
\label{decc}
\end{equation}
During this deceleration phase, the first term within parenthesis
is largely dominant, and the second term can be considered to be
a constant where $r^2 \dot{r}^2$ is approximated by its mean 
value $< r^2 \dot{r}^2 >$.
Hence, we have $\dot{r}=\dot{r_1} e^{-\alpha (r-r_1)}$.

This approximation is plotted on Fig.~\ref{velocity} with long-dashed
lines. This is valid as long as the velocity is large enough for the drag
to dominate the radiation pressure.
Subsequently, the radial velocity significantly decreases
and the drag decreases to the radiation pressure's level.
Thus the deceleration takes place when $r_1<r<r_2$, 
where $r_2$ is defined by
\begin{equation}
\dot{r_1} e^{-\alpha (r_2-r_1)}=\sqrt{\frac{GM}{C\rho}} r_2^{-1}.
\end{equation}

\subsection{Stationary outflow}
\label{Stationary outflow phase}

When $r>r_2$, then we simply have $d^2r/dt^2 \sim 0$: this is the stationary
outflow where the gas drag exactly compensates the radiation pressure.

From Eq.~\ref{eq2}, we get $\dot{r}= \sqrt{\frac{GM}{C\rho}} r^{-1}$.
This gives $n(r)\propto r^{-1}$, and $F(r)\propto r^{-2}$,
exactly as observed around BD+31$\degr$643.

This is represented on Fig.~\ref{velocity} by the short-dashed
lines. \\

When the particle velocity becomes very small, the thermal term
is dominant in the gas drag (Eq.~\ref{Fg}). 
This situation occurs when $r>r_3$ where 
$r_3\equiv \sqrt{\frac{GM}{C\rho}} v_T^{-1}$.
Eq.~\ref{eq2} is then 
\begin{equation}
\frac{d^2r}{dt^2}   \approx \beta \frac{GM}{r^2} -C'\beta\rho(\frac{dr}{dt})
\label{eq2them}
\end{equation}
where 
$C'= 375 m_\mathrm{H} \mu v_T
     \approx 1.3\cdot 10^{-9}\sqrt{(T/30\mathrm{K})}$cm$^3$year$^{-1}$.
During this phase, we still have $d^2r/dt^2 \sim 0$, which gives
$\dot{r}= \frac{GM}{C'\rho} r^{-2}$, and $F(r)\propto r^{-1}$.
This approximation is plotted on Fig.~\ref{velocity} with dotted
lines.

\section{Application}
\label{appli. The bd disk}

\begin{figure}
\resizebox{\hsize}{!}{\includegraphics{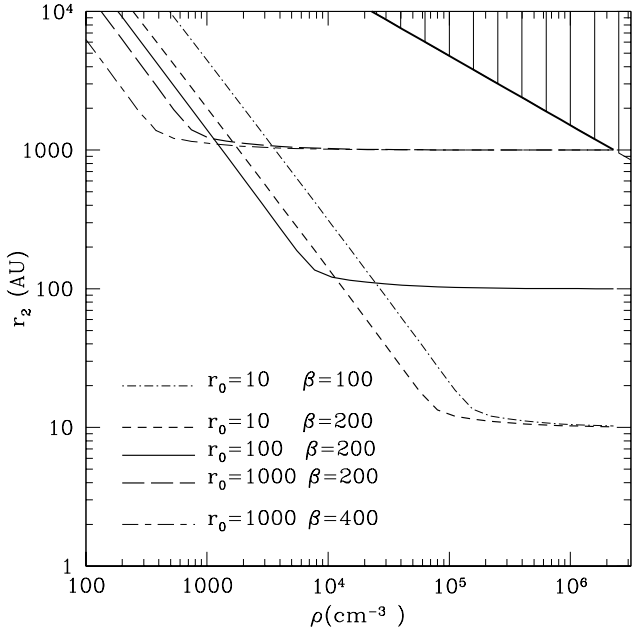}}
\caption[]{Plot of the radial distance $r_2$, as defined in the
text. The hashed zone represents the region where $r>r_3$, for a temperature
$T=30$~K. 
Above the lines representing $r_2$ and below the hashed zone, 
we have $\dot{r} \propto r^{-1}$, hence $F(r)\propto r^{-2}$ as 
observed around BD+31$\degr$643\ between 2300 and 6000~AU.
We see that $r_2<2300$~AU gives $\rho \ge 10^3$cm$^{-1}$,
and $r_3>6000$~AU gives $\rho \le 10^5$cm$^{-1}$, 
}
\label{r2r3}
\end{figure}

\subsection{The radial velocity versus gas density}

We applied the previous calculations and approximations to the
case where a grain is produced at a distance $r_0=100$~AU from the star and
with $\beta=200$.
We have plotted the dust radial velocity ($dr/dt$) as a function of the 
distance
to the star ($r$) for different gas density ($\rho$) (Fig.~\ref{velocity}).
The full Eq.~\ref{eq1} has been numerically solved, and the result is
given by the solid line in Fig.~\ref{velocity}.

The fit to the different phases: acceleration (dot-dashed lines), deceleration
(long-dashed line), stationary outflow (short-dashed and dotted lines)
are calculated with the approximation given by the equations of
Sect~\ref{Outflow with gas drag}. The 
superposition of the exact solution of Eq.~\ref{eq1} and the 
different fits validates these approximations and the corresponding 
domain of application given in the previous section.
This also validates the different positions where
the changes of regime occur ($r_1$, $r_2$, $r_3$). 

For $\rho = 30 $cm$^{-3}$, we see that 
the particle is accelerated between 100 and
1000~AU; it is then decelerated between 1000 and $10^5$~AU and reaches the
stationary outflow beyond that point.

For $\rho = 300 $cm$^{-3}$, the acceleration stops at 300~AU, and
the deceleration takes places between 300 and 5000~AU; the stationary outflow 
starts at $r_2=5000$~AU. The distribution changes again at $r_3=10^5$~AU when 
the thermal drag becomes dominant. 
This gas density is too low to explain the observed distribution
of dust around BD+31$\degr$643. 

For $\rho = 3000 $cm$^{-3}$, the deceleration phase is very short
(between $r_1=150$~AU and $r_2=400$~AU). 
The stationary outflow between 400~AU and ($r_3=$)30\,000~AU gives a
radial velocity $\dot{r} \propto r^{-1}$. 
Then the disk surface brightness distribution is $F(r)\propto r^{-2}$,
exactly as observed around BD+31$\degr$643. The distribution changes at 30\,000~AU 
beyond which we have 
$\dot{r} \propto r^{-2}$, and consequently $F(r)\propto r^{-1}$.

For $\rho = 300\,000 $cm$^{-3}$, there is no deceleration phase at all.
The thermal drag is dominant when $r>r_3\cong 2800$~AU. This gas density
is too large to explain the observed distribution of dust around 
BD+31$\degr$643.\\

\subsection{The BD+31$\degr$643\ disk}

In the BD+31$\degr$643\ candidate disk, the distribution observed between 
2300 and 6000~AU can be explained
if the outflow is stationary between $r_2$ and $r_3$ as described in 
Sect.~\ref{Stationary outflow phase}.
This happens only if $r_2<2300$~AU and $r_3>6000$~AU. 
In Fig.~\ref{r2r3} we have plotted these distances for different
ratio $\beta$ and initial distance $r_0$.
We can see that these conditions are satisfied if we have
$10^3$cm$^{-1}\le \rho \le 10^5$cm$^{-1}$.

Incidentally, another solution could be found if the dust is in the 
stationary outflow in the thermal drag regime ($r>r_3$) and 
with a radially decreasing gas density $\rho \propto r^{-1}$. 
In that case, we have $\dot{r}\propto r^{-2}/\rho\propto r^{-1}$, and 
again $F\propto r^{-2}$.
But this would imply that the gas density should 
be larger than $10^6$~cm$^{-3}$ which is incompatible with 
the observed gaseous column density.

\section{Numerical results}
\label{Numerical results}

\begin{figure}
\resizebox{\hsize}{!}{\includegraphics{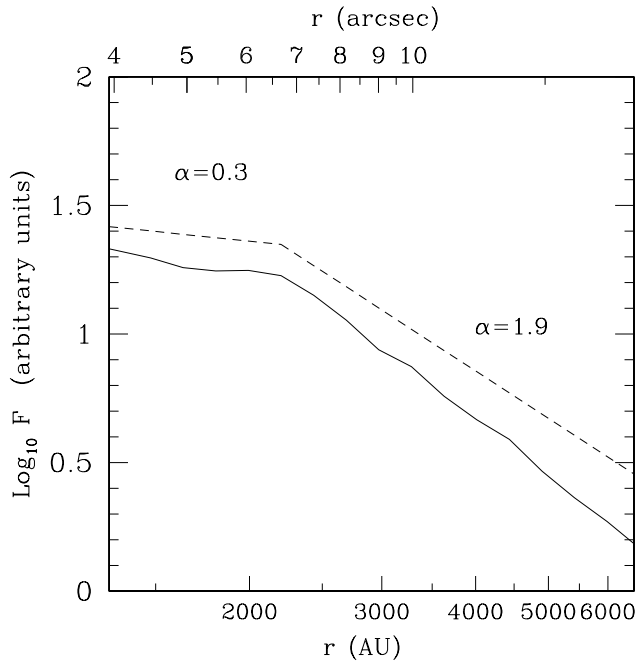}}
\caption[]{Brightness distribution of a dust disk seen edge-on 
and produced by parent bodies on circular orbits between 2000 and 2300~AU.
The gas drag is taken into account with $\rho=10^{4}$cm$^{-3}$.
The particles have a $\beta$ ratio of the gravitational forces to 
radiation forces between 10 and 1000.
This distribution is very close to the surface brightness 
distribution observed around BD+31$\degr$643\ by Kalas \& Jewitt (1997)
(dashed line, slightly arbitrarily shifted to be better compared.)}
\label{profil}
\end{figure}

For a mixture of particle size (and $\beta$), a radial
distribution of the sources and the exact treatment of the
equation of motion (Eq.~\ref{eq1}), numerical simulations are required.
We built a code to solve this equation and to calculate
the distribution of the dust particles in the steady state of the
equilibrium between 
the continuous replenishment by a distribution of parent bodies and 
the ejection by the radiation pressure.

A wide range of parameters for the dust size distribution, 
the gaseous density and velocity ($\vec{v_g}$), and the 
initial parent bodies positions have been used.
The simulations show that dust produced within 2300~AU is
able to explain the surface brightness of the disk, 
provided that the gas density
around the star is $\rho\sim 10^4$~cm$^{-3}$.
An example of such a simulation is given in Fig.~\ref{profil},
with a size distribution 
corresponding to $\beta\in [10,1000]$.
Any size distribution gives the same results, if particles
with $\beta\la 10$ are in negligible quantity.

All these parameters are very consistent with the observations 
of the large albedo of grains in the blue and 
the estimated density of gas at the center of the cluster IC~348
(Sect~\ref{The bd disk}).
The only parameter newly constrained by these simulations
is the position of the parent bodies, which 
must be located between 2000 and 2300~AU from the star
to explain the observed decrease of slope within 2300~AU.
It must be mentioned here that BD+31$\degr$643\ is 200~times 
brighter than $\beta\:$Pictoris,
so that the distance of evaporation of CO (150~AU around $\beta\:$Pictoris) 
corresponds to about 2100~AU around BD+31$\degr$643.
But this change of slope can also be linked to the limited resolution
of the observations and needs confirmation.

It must also be noted that the velocity of the star relative to the ambient 
gas ($\vec{v_*}$) could modify the appearance of the disk. However, as can be
inferred from the Fig.~\ref{velocity}, simulations show that asymmetries
would be visible only for distances 
$r\ga 10\,000\ (v_*/1{\rm km\ s}^{-1})^{-1}$AU.
The observations are thus consistent with the present modeling if 
$v_*\la 2 {\rm km\ s}^{-1} $. Considering that the star is at the 
center of the cluster and that only the transverse motion
would give noticeable asymmetries, this last condition 
is certainly not unlikely.

Finally, the gas drag is the dominant phenomenon which allows
to explain the dust radial distribution, but this also rules out the
possibility to better constrain the dust origin.

\section{Discussion}

The dynamics of particles in the BD+31$\degr$643\ disk and the observed 
distribution of dust can be explained by the simple model of dust outflow
presented here. 

Unfortunately, the issue of the origin of the very small particle size 
remains unexplained. It is not clear why we see so much small particles
around BD+31$\degr$643, whereas the dust 
around other main sequence stars with known infrared excess is usually
much larger (e.g., Habing et al. 1996).

Also the process which produces these grains is still not determined. 
The present model of the BD+31$\degr$643\ disk does not discriminate 
between production by collision or by evaporation.
However, a change of
slope and possible inner hole are observed at a distance of 2000~AU,
corresponding to the sublimation distance of CO ($\sim$ 2100~AU).
Production of dust by evaporation could start if some bodies
were heated by the central star after the dissipation of the
dense optically thick disk from which a planetary system formed.
Alternatively, parent bodies could evaporate 
if they were scattered planetesimals on eccentric orbits in a
forming Oort-like cloud or in a Kuiper-like outer belt
(Scattered Icy Objects? Duncan \& Levison 1997).

Finally, the particles around 
BD+31$\degr$643\ can have a similar origin and nature
as the dust grains around $\beta\:$Pictoris, but both disks
present different characteristics. The main differences are
related to the large distance ratio
compensated in part by the larger luminosity of BD+31$\degr$643, and
consequently the dynamics of dust here expelled by the strong radiation
pressure.

\begin{acknowledgements} 
We are particularly indebted to an anonymous referee for his very useful 
comments.
We would like to express our gratitude to S. Loi\-seau
for his critical reading of the manuscript. 
We also warmly thank A. Dutrey for very fruitful discussions.
\end{acknowledgements}

\end{document}